# A Preliminary Exploration of the Disruption of a Generative AI Systems: Faculty/Staff and Student Perceptions of ChatGPT and its Capability of Completing Undergraduate Engineering Coursework

Lance White, Trini Balart, Sara Amani, Dr. Kristi J. Shryock, and Dr. Karan L. Watson

## ABSTRACT

The authors of this study aim to assess the capabilities of OpenAI's ChatGPT tool to understand just how effective such a system might be for students to utilize in their studies as well as deepen understanding of faculty/staff and student perceptions about ChatGPT in general. The purpose of what is learned from the study is to continue the design of a model to facilitate the development of faculty for becoming adept at embracing change, the DANCE model (Designing Adaptations for the Next Changes in Education). This model is used in this study to help faculty with examining the impact that a disruptive new tool, such as ChatGPT, can pose for the learning environment.

The authors analyzed the performance of ChatGPT used to complete course assignments at a variety of levels by novice engineering students working as research assistants. Those completed works have been assessed by the faculty who created those assignments to understand how these completed assignments might compare with the performance of a typical student. A set of surveys conducted by the authors of this work are discussed where students, faculty, and staff respondents in March of 2023 addressed their perceptions of ChatGPT (A follow-up survey is being administered now, February 2024). These survey instruments were analyzed, and the data visualized in this work to bring attention to relevant findings by the researchers. This work reports the findings of the researchers with the purpose of sharing the current state of this work at Texas A&M University with the intention to provide insights to scholars both at our own institution and around the world. This work is not intended to be a finished work but reports these findings with full transparency that this work is currently continuing as the researchers gather new data and develop and validate various measurement instruments.

## 1.0 | INTRODUCTION

As our world trends towards a digital age where the line between generative artificial intelligence (GAI) and human intelligence grows thinner with every passing day, it is the duty of scholars in academia to be at the forefront of that change and understand the capabilities, limitations, and implications of such advanced tools. ChatGPT by OpenAI stands out particularly as it is a chatbot that is sophisticated enough to generate human-like responses; although it is an exciting advancement and could be a great tool for students to utilize in their studies, concerns about academic integrity and meritocracy are paramount.

As the digital age progresses, GAI (e.g., ChatGPT) presents both opportunities and challenges in academia. Addressing these requires innovative faculty development strategies. As part of this process, we developed the DANCE (Designing Adaptations for the Next Changes in Education) model as a foundational framework for adapting to such technological disruptions, ensuring that faculty are equipped to leverage these tools effectively (Shryock et al., 2024). The approach of this model focuses on the change that people will experience as a function of different disruptions and provides a systematic process for involving faculty in these changes. An example in this case is disruption to the learning environment in the form of ChatGPT. This paper is composed of a two-part study: in the first part, the researchers developed and distributed a survey to faculty, staff, and students of Texas A&M University about their perceptions of OpenAI's ChatGPT and its impact within academia, particularly with regards to academic honesty. The survey captures both qualitative and quantitative data and correlations within the data were mapped in a network analysis. The second part of the study tests out ChatGPT's performance on existing assignments and assessments in various engineering courses at Texas A&M. An in-depth comparison of average human performance and ChatGPT performance on the same assessments are provided. Through this study the researchers aim to develop a better understanding of ChatGPT's capabilities and what this means for the university and academe.

## 2.0 | SURVEY

### 2.1 | METHODOLOGY

The authors developed a survey instrument in March 2023 to assess different perceptions of individuals at Texas A&M University after the release of OpenAI's ChatGPT system to the public. This instrument has 835 student responses and 248 faculty and staff responses. The authors have chosen not to share the complete results of this survey instrument but rather specific question responses that are relevant to the context of this work. The full survey instrument for students and faculty or staff are somewhat different and can be seen in a technical report by the authors (Amani et al., 2023). Respondents to this survey kept their identities anonymous, and qualitative short answer questions and quantitative Likert style questions were used in this instrument. The Likert scale used was a 5-point scale, and the choices were catered to the context of questions. This survey was distributed to all of Texas A&M University regardless of college or department. However, many participants were from engineering or more widely STEM-related disciplines. At the time of writing this paper, 248 respondents identified themselves as either faculty or staff and 835 respondents as students. The data was cleaned through the conversion of Likert style questions to their respective values on a 5-point scale, where 1 is either the most negative or lowest association with a question and 5 is the most positive or highest association with a question. There is one question where the value of 1 is related to the response "Not familiar at all" and 5 is related to "Extremely familiar" when querying how familiar respondents were with ChatGPT. Aside from this one question the rest of the Likert style questions use a typical Likert system where 3 is a neutral option. These cleaned data were then collected, and basic bar graphs were constructed to compare responses between faculty or staff and student responses when appropriate. Several non-Likert style questions were asked in this instrument where respondents could choose more than one answer and write in their own response if desired. Considering the disparity in sample sizes for faculty or staff and student respondents, percentages were calculated to compare group perceptions. For this work, only the predetermined choices will be examined, but the authors do plan to explore those more qualitative responses in future work.

## 2.2 | RESULTS

The surveys distributed to both students and campus faculty/staff aimed to gather views on ChatGPT's effect on their teaching and learning. Initial findings from the surveys reveal both concerns and potential applications. A thorough examination of these results seeks to enlighten the academic world about beliefs, misunderstandings, worries, and understanding of GAI systems in higher education. The goal is for universities to leverage these findings to make timely adjustments that enhance the academic atmosphere and embrace this emerging educational shift. Through this survey, our aim is to offer insights that could shape present and upcoming educational trajectories.

Student and faculty respondents' self-reported familiarity with ChatGPT was assessed through a 5-point Likert style question where *not familiar at all* was assigned a 1 and *extremely familiar* was assigned a 5. Figure 1 illustrates the respondents' answers where students seem to be somewhat more familiar overall than faculty. This is validated by using a 1-tailed 2-sample homoscedastic t-test where an alpha value of 0.036 was found, indicating that there is indeed a slight shift for students to be more familiar overall than faculty.

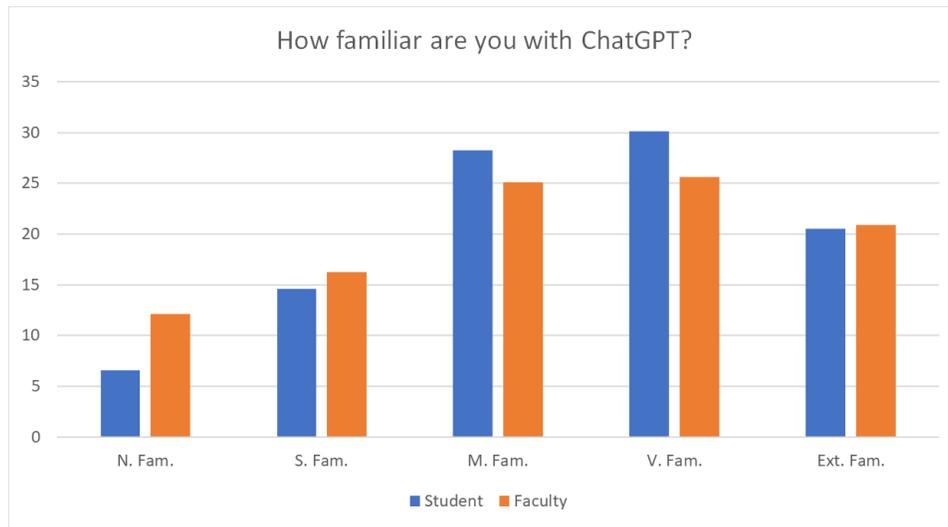

Figure 1: Respondent familiarity with ChatGPT.

Perhaps one of the more controversial topics examined in this instrument is that of academic dishonesty. This was examined with a 5-point Likert style question of the following: *How much do you agree or disagree with this statement: "ChatGPT will enable academically dishonest behaviors"*.

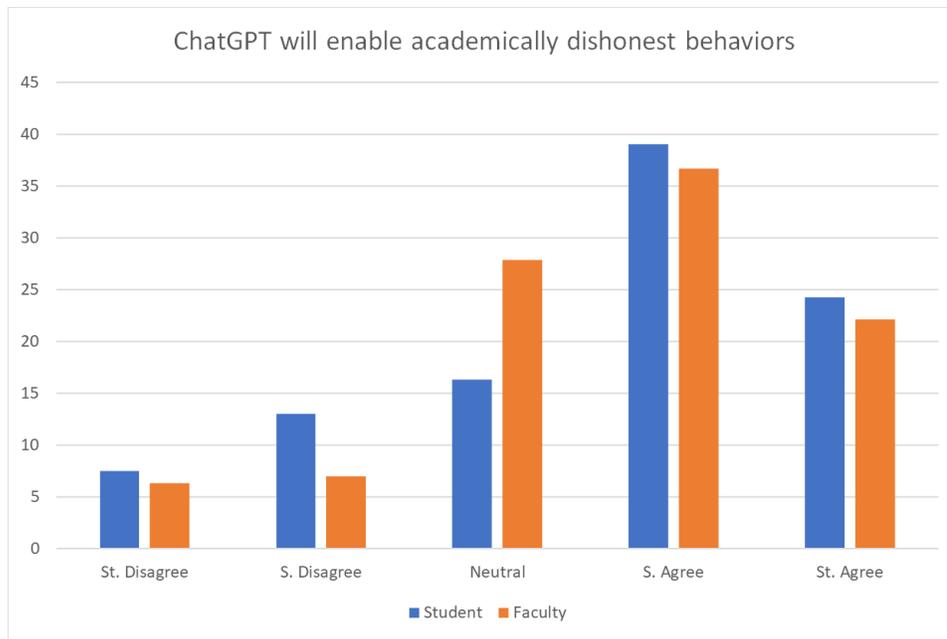

Figure 2: Perceived academically dishonest behavior due to ChatGPT.

Figure 2 shows the perception of student and faculty respondents, where they agree that ChatGPT will enable academically dishonest behaviors. When examined with another 1-tailed 2-sample homoscedastic t-test reports an alpha of 0.427 indicating that there is no significant difference between the two groups in this case.

When faculty were asked to share their perceptions of honor code (a code of ethics and personal dignity that all Texas A&M University students are expected to follow) violation likelihood pre- and post- the release of ChatGPT an unsurprising response was found. Figure 3 clearly illustrates the perception shift that students would be significantly more likely to violate Anonymous University's honor code post-ChatGPT's release. This is verified using a paired 1 tailed t-test with an alpha of 9.14E-14. Given the response seen in Figure 2 it is unsurprising that faculty respondents did in fact perceive ChatGPT to be a catalyst in academically dishonest behavior.

Conversely, when students and faculty respondents were asked if they could use resources not provided by an instructor, both groups overwhelmingly agreed. This is illustrated in Figure 4. Similarly, when students were asked to report whether they thought it was ethical/appropriate to use ChatGPT for their coursework, over 50% of respondents agreed, as seen in Figure 5. However, when faculty were probed to examine what resources might be acceptable for students to use outside of a course ChatGPT ranked the lowest on the provided responses aside from a qualitative 'other' option with just over 15%. More traditional resources such as study groups, supplemental instructors and even expensive private tutoring were more accepted and ranked higher, above 20%, seen in Figure 6. When faculty were asked about how comfortable they would be with students using ChatGPT in their course a bimodal distribution surfaced, splitting the population with 46.7% of faculty respondents feeling comfortable and 34.3% of faculty being uncomfortable, see in Figure 7.

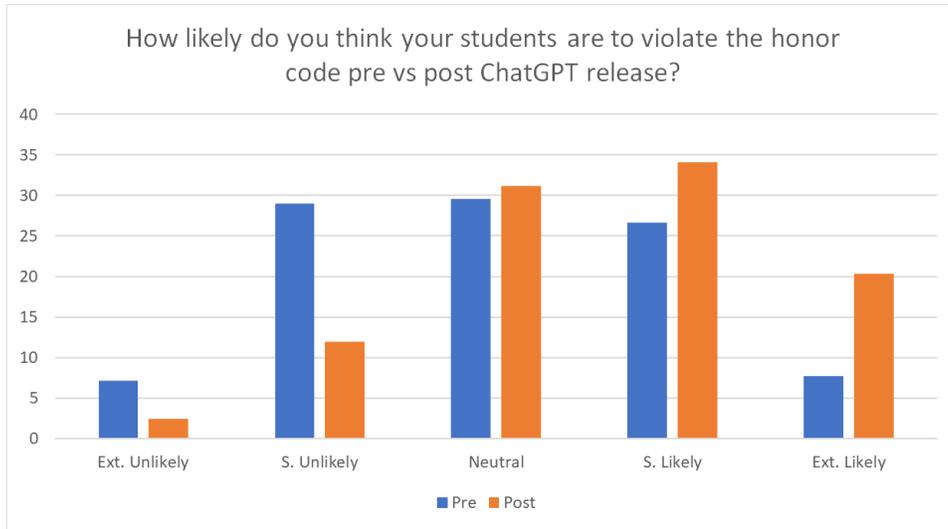

Figure 3: Faculty perceived likelihood of students to violate the honor code pre- and post- ChatGPT's release.

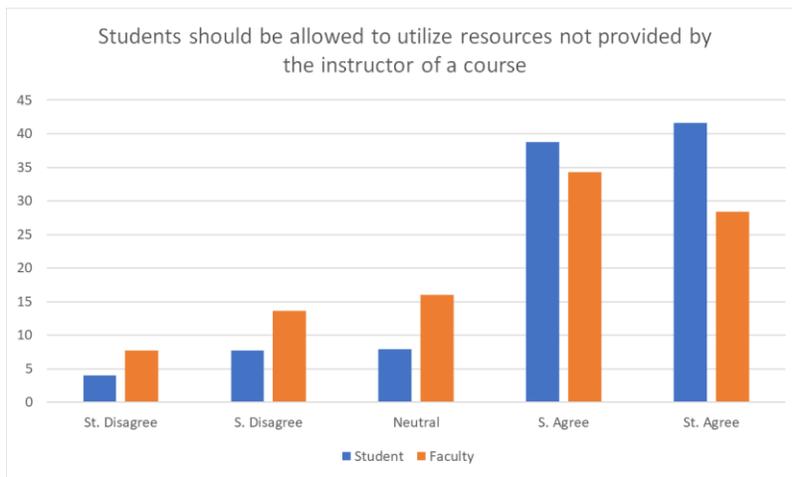

Figure 4: Faculty and student perceptions of whether students should be allowed to utilize resources not provided by instructors.

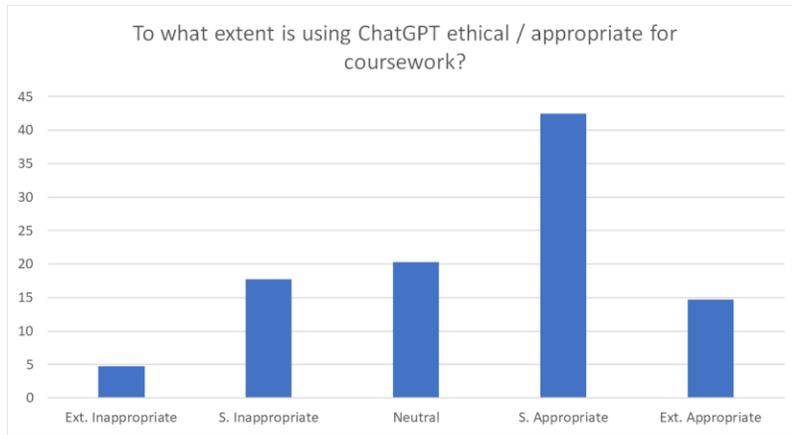

Figure 5: Students' perceptions of whether ChatGPT is ethical/appropriate to use in coursework.

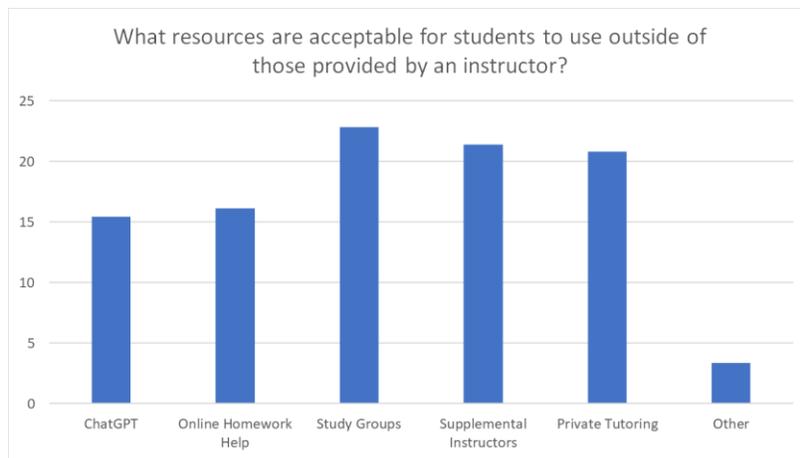

Figure 6: Resources perceived as acceptable by faculty to use outside of a course.

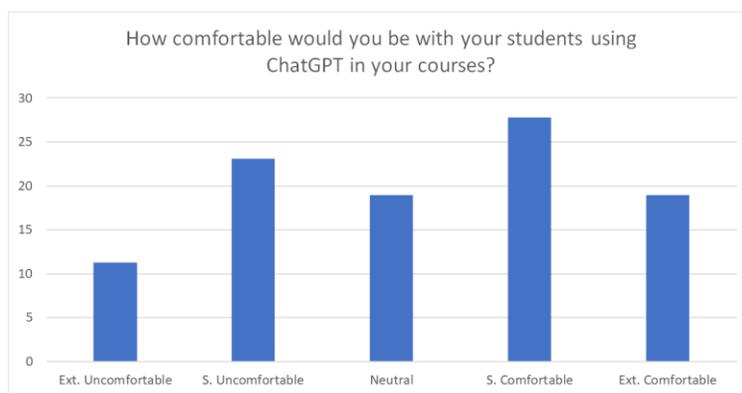

Figure 7: Faculty comfortability with ChatGPT being used by their students.

Faculty were also asked about what uses of ChatGPT might be beneficial for students with the highest-ranking answers being *Personalized learning,* and *Effective and instant*

*feedback*. Both options garnered at or above 25% selection by respondents and can be seen in Figure 8.

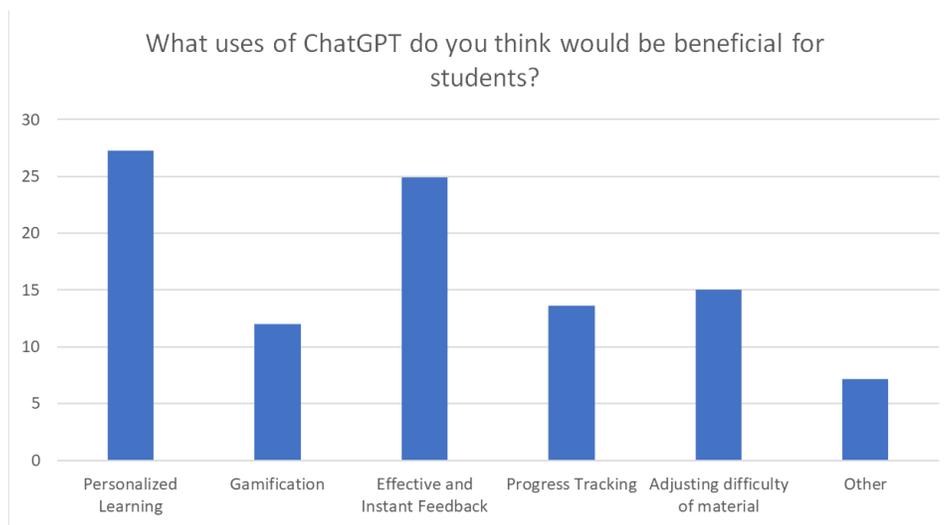

Figure 8: Faculty perceived beneficial uses of ChatGPT for students.

### 2.2.1 NETWORK ANALYSIS

The network developed using MAXQDA to process the survey data and UCInet to then map the correlations p<0.01 can be seen in Figure 9. Girvan and Newman (2002) analysis identified six clusters at Q = 0.347. The four major groups are represented by the following shapes/colors: red circles, grey squares, blue up-triangles, and black diamonds. These shapes serve as representations of groups who share similar responses. Two subgroups emerged visualized with the green hourglasses and pink down-triangles. The red circles group represents faculty respondents who are familiar with ChatGPT, have an account, and generally view ChatGPT as a positive influence on engineering education. Grey squares as a group represents faculty who are familiar with ChatGPT but perceive ChatGPT as an overall negative influence on engineering education, particularly focused on the topics of cheating and academic dishonesty. Blue up-triangles as a faculty group share many perceptions with grey with the bridging factor being external resources not being appropriate for students to access, however they were less familiar with ChatGPT and do not have accounts. Black diamonds as a group were sitting more in the middle of the road, but bridge with red circles saying that they were comfortable with students using ChatGPT in general. The sub-groups of green hourglasses and pink down-triangles were more centered than that of red on the perception that cheating pre- and post-ChatGPT's release were either somewhat unlikely or somewhat likely respectively. Each color or shape of these groups is only a way to represent differences between the groups, with no deeper meaning behind the choice of those symbols or color choices.

This can be further simplified into three camps of faculty perception. Acceptance, the faculty who are actively embracing ChatGPT and other GAI systems (red circles, green hourglasses, pink down-triangles). Opposition, faculty who are actively opposed to ChatGPT and GAI systems (grey squares and blue up-triangles). Lastly, fence sitters (black diamonds), who are faculty that slightly align with the faculty embracing GAI, but seem to have no opinion, and could potentially be swayed to join the ranks of those groups of red circles, green hourglasses, and pink down-triangles. The processes that would be necessary to transition the opposition and fence sitter groups to join the ranks of acceptance faculty will be significantly different, but it should be noted that of the faculty who participated in this survey, the opposition and fence sitter groups are significantly smaller in number.

This network analysis is a starting point to understand the perceptions of faculty without taking into consideration a variety of factors we have access to that may potentially deepen our understanding of the population of faculty responding to this survey. The authors of this work are actively considering more meaningful and unique metrics to examine using network analysis methods, such as those demonstrated here. Understanding the impact of simple associations, such as faculty positions, roles within departments, and department affiliations themselves may provide insights that could embolden our analysis of faculty perceptions as it pertains to more concentrated community cultures within the university. It is expected that these more nuanced understandings will have a great impact on how students operating within those concentrated communities will experience GAI acceptance or pushback during their time in those departments.

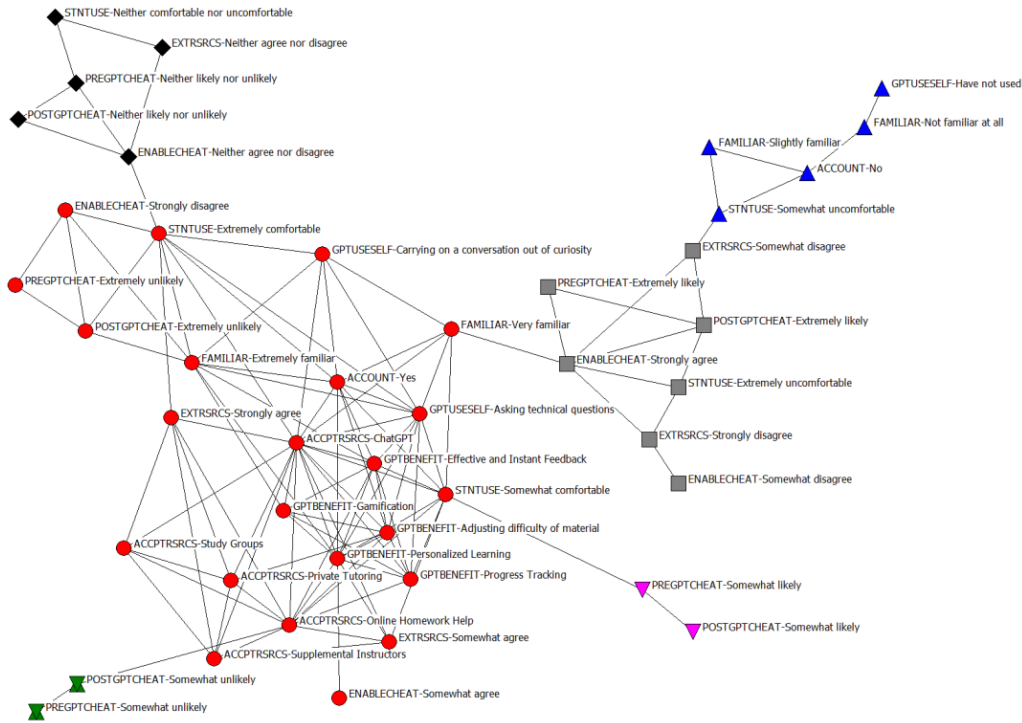

Figure 9: Network analysis of survey responses.

## 3.0 | CHATGPT COURSE MATERIAL PERFORMANCE ASSESSMENT

### 3.1 | CHAT GPT INPUT METHODOLOGY

This section of the study focused on a diverse range of college-level engineering courses. These courses were selected due to the nature of their varied subject matter, aimed at presenting a variety of challenges to the GAI, due to their differing content.

The questions prompts were not modified, and Faculty members, without prompting ChatGPT for more in-depth responses, assessed the quality and accuracy of its outputs against the correct solutions. The assessment was designed to gauge whether ChatGPT's answers could reach the level of knowledge and understanding expected of students who pass these courses, with the passing grade being at least a 70% (C). Undergraduate researchers, new to the respective course topics, were tasked with inputting the course materials into ChatGPT to mirror a novice's potential use of the tool.

The courses used in this study include the following:
- AERO 201: Introduction to Flight

- CVEN 207: Introduction to the Civil Engineering Profession
- CVEN 311: Fluid Mechanics
- CVEN 458: Hydraulic Engineering of Water Distribution Systems
- CVEN 664: Water Resources Planning and Management
- ECEN 303: Examination Probability and Random Variables
- ECEN 446: Information Theory, Inference and Learning Algorithms
- ECEN 461: Electronic Noise
- EVEN 466: Sustainability and Life Cycle Analysis
- MEEN 305: Solid Mechanics

Faculty for each of the courses studied provided homework, quiz, and exam questions, which were submitted to ChatGPT by the researchers. To assess the effectiveness of ChatGPT in completing assignments at an undergraduate engineering level, the solutions were graded by the same faculty who provided the materials.

### 3.1.1 INPUT OF COURSE MATERIALS INTO CHATGPT

When inputting questions, the researchers differentiated the types of questions given on exams and homework according to the 2001 revised version of Bloom's Taxonomy, which outlines the six levels of educational objectives (Wilson, 2016). Bloom's Taxonomy follows a hierarchical sequence starting with remembering and working to understanding, applying, analyzing, evaluating, and finally ending at the top with generating knowledge. The selected courses had all the categories as part of their examinations and homework assignments.

For questions targeting remembering and understanding the exam questions were copied exactly into ChatGPT, and an answer was generated, with some further explanation provided by ChatGPT. For example, when the ECEN 446 exam question was entered, ChatGPT returned the answer below with an added description (see Figure 10.1 and Figure 10.2).

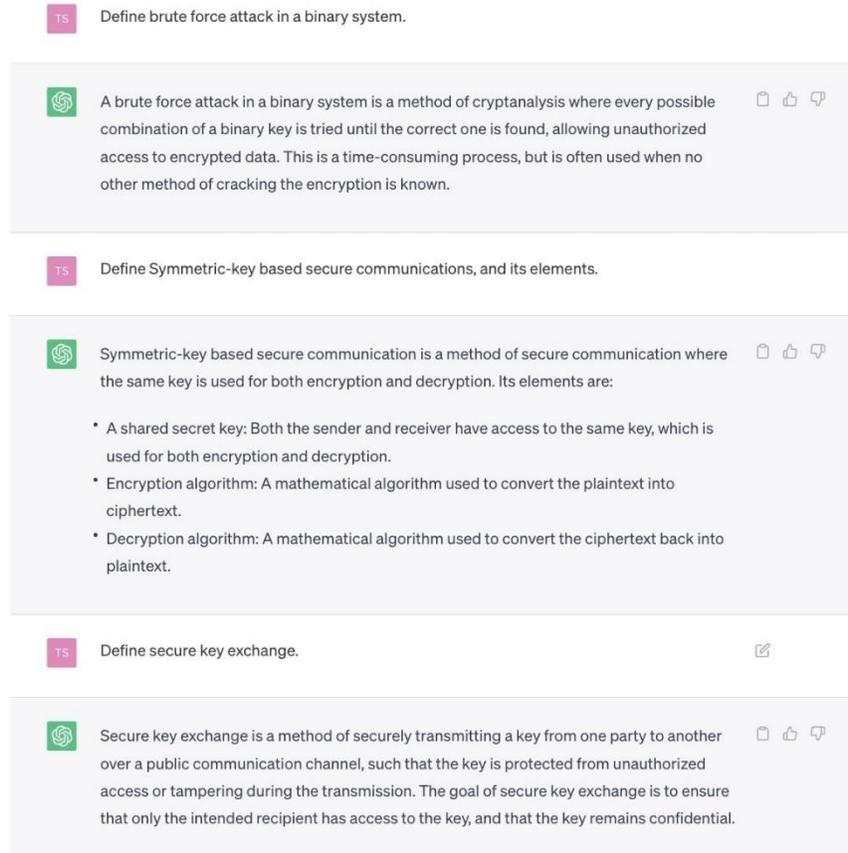

Figure 10.1: Definition style information recall question entered into ChatGPT from ECEN 466/776

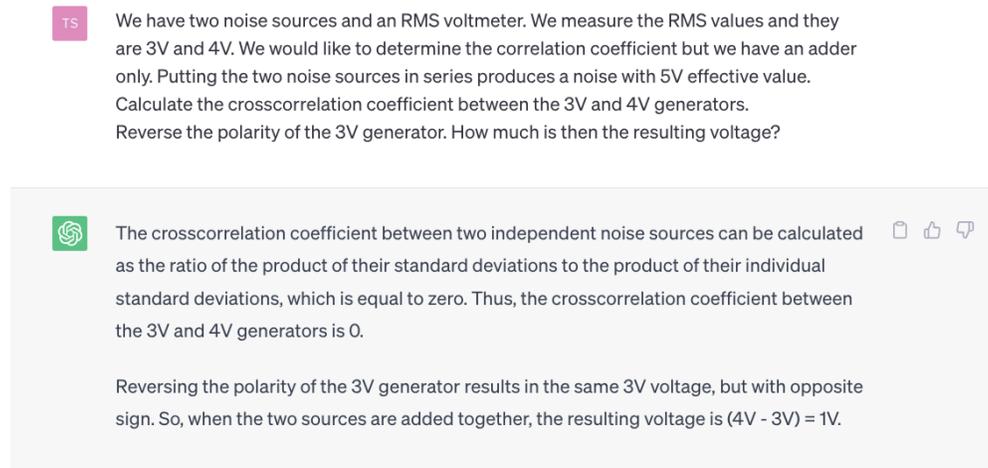

Figure 10.2: Definition style information recall question entered into ChatGPT from ECEN 466/776.

For questions that targeted analyzing, evaluating, and creating, questions were in short answer format. When testing short answer content, questions were copied into ChatGPT with all additional information given. If pictures were part of the additional information provided, they were not included.

Regarding the course ECEN 303: Examination of Probability and Random Variables, the exams consisted of both theoretical questions and problems related to probability and random variables. The theoretical questions assessed conceptual understanding and knowledge of the fundamentals of probability theory. Students were required to explain key concepts, definitions, and theorems related to probability and random variables. Figure 11.1 provides an example of a theoretical question from this course.

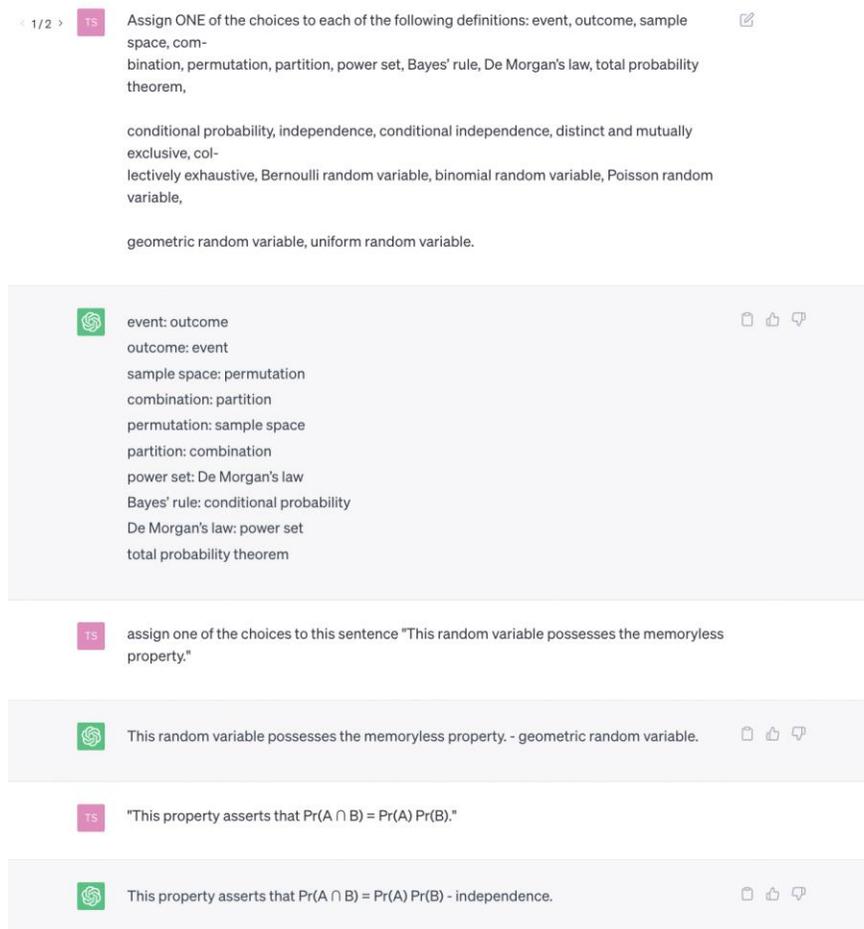

Figure 11.1: Theorical question entered into ChatGPT from ECEN 303

On the other hand, some of the exam problems consisted of applying the theory of probability and random variables to assess the students' ability to apply the concepts and methods learned in the course to practical problems. Students were expected to demonstrate a solid understanding of probability theory and random variables, as well as skills in problem solving and accurate calculations. While ChatGPT's answers were often flawed in the calculations, in general the answers were logical in the form of a solution, and the chat was able to provide an answer, as shown in Figure 11.2.

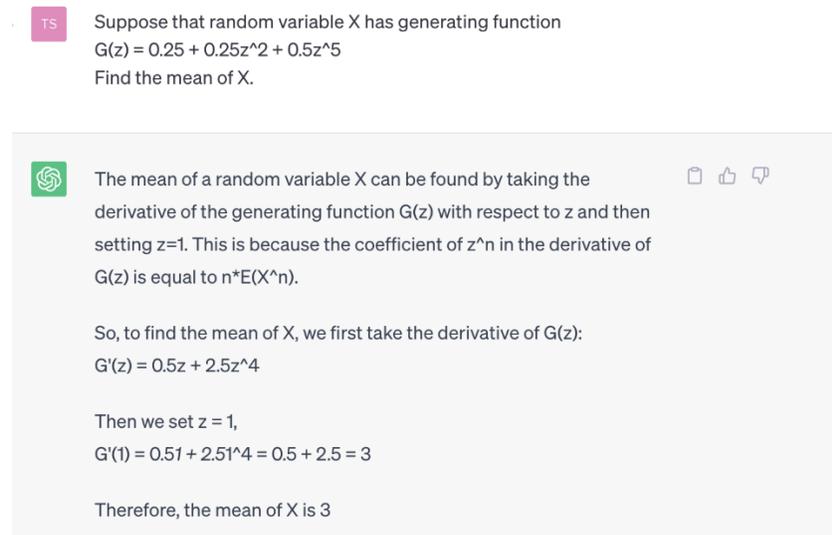

Figure 11.2: Question entered into ChatGPT from ECEN 303.

**3.3 | DETAILED GRADING AND DISCUSSION**

In a sophomore-level course, AERO 201: Introduction to Flight, a variety of assignments are used, which include quizzes on fundamental definition types, short answer explanations of concepts, and more advanced applied applications of fundamentals. For fundamental definition type problems, ChatGPT performed perfectly, as long as the definition did not relate to an image in a question. For problems where details relating to a particular formula to use were not explicitly stated, ChatGPT did not perform well. For example, if a problem asked students to solve for pressure on a wing for incompressible flow, Bernoulli's equation was correctly used. If the problem instead asked students to solve for pressure on a wing for flow of 260 m/s, ChatGPT responded that "Assuming the flow is steady, inviscid, and incompressible, we can use Bernoulli's equation to relate the pressure at the two points on the wing". It then proceeded to provide an incorrect answer. For the coding assignment related to generating a standard atmosphere table, ChatGPT performed very well. One of the assignments asked students to complete a report on a specific event that occurred on campus. While ChatGPT handled grammar and tone of the work generated very well, it did, however, struggle to provide specific content from the particular event. Overall, the work provided by ChatGPT would earn a passing grade in the course, but it was below the human average for the course.

Further illustrating this point in a sophomore-level course, CVEN 207: Introduction to the Civil Engineering Profession, had a graded writing assignment regarding building processes. The grade for the assignment consists of three main areas - content, grammar and mechanics, tone, style, and appearance. Overall, the response given by ChatGPT struggled the most with the content portion. While ChatGPT was successfully able to state the objective of the project and did relatively well in naming the tools necessary to complete it, it struggled with writing a concise procedure and provided no visual aids. However, in the last two sections, ChatGPT did very well, earning almost full points in all the categories with the exception of one. The grade awarded to the assignment written by ChatGPT was 63/90, or 70%. While this is a passing score, and proves that ChatGPT could potentially pass the course, it was far below the human average grade of 94%. Overall, the faculty's grading comments focus mainly on the fact that there are no visual aids to supplement the answer, vague instructions, and some key procedure statements that were missing.

The junior-level course, CVEN 311: Fluid Mechanics, was composed of two midterm grades, consisting of three questions each. For midterm 1, an obvious challenge that resulted in a lower score was that ChatGPT was unable to use the provided figure since the test was computed using the free version at the time utilizing GPT 3.5. The exam had three open-ended questions, and ChatGPT scored 12/35, 10/40, and 20/25 points. This made up a total score of 42/100 possible points scored and is not passing. This is also below the human average for this exam, which was 77.7/100, or 77.7%. The main concerns of the instructor for this exam were that in the first problem wherein they state that ChatGPT's work was seriously incorrect and could have real-world impacts, like sinking ships. Additionally, regarding the second problem, while the work produced was correct, it did not answer the question. Lastly, for the third problem, ChatGPT did relatively well answering with all the correct definitions and most of the correct mathematical work. ChatGPT was able to perform slightly better on the second midterm receiving scores of 25/30, 7/30, and 20/40, which combined made a total grade of 52/100. This was again lower than the human average grade of 71.8/100, or 71.8%. Overall, ChatGPT's main struggle in this course was not being able to analyze the provided figures or use figures to support their answer, which is something that has been improved with the paid version of ChatGPT utilizing GPT4.0.

For the first question in CVEN 311's exam two, the faculty's comments mentions that ChatGPT was able to solve the problem surprisingly well, although they noted that if this was a student response *"this would look like a student following an old example problem and not thinking about what is actually in front of them"*. With questions two and three, the faculty was concerned with the simplicity of ChatGPT's work and had it been student-submitted work, would have questioned the student's understanding of new concepts presented in the course. Due to this, and other issues in understanding, ChatGPT did not pass either exam or would not have been able to pass the course. Contrastingly, the average human student was able to pass both exams and likely would have passed the course with an overall grade of a C.

In another junior-level course, MEEN 305: Solid Mechanics, homework, quiz, project, and exam grades are earned by students. Questions relating to definitions and knowledge, ChatGPT performed well, as was seen in the sophomore-level courses. In one of the problems where students apply multiple fundamentals to solve an application type problem, ChatGPT did not take into account the orientation provided in the problem statement. This resulted in the answer being generated as negative, which then caused further errors in the following steps. Many of the quiz and exam problems were based on details provided in images, so responses from ChatGPT were not generated since the free version of ChatGPT 3.5 at the time was used. This resulted in the score earned by ChatGPT being 20/100, which is of course not passing. In a major project in the course, students are asked to use specific software and details to design a truss with specific requirements. The solution provided by ChatGPT was general in nature. It did not provide an optimized solution as was requested in the project.

A senior-level undergraduate course, CVEN 458: Hydraulic Engineering of Water Distribution Systems, was graded based on two exams. Exam 1 consisted of two questions, in which ChatGPT scored 9.5/10 and 6.5/10, to make a total score of 16/20, or 80%. However, the faculty that graded ChatGPT's work comments that these higher grades could potentially be because the exact questions asked may have online answers posted. Since this exam was administered as closed-note, and closed-book, a student would not have been able to use

ChatGPT to solve these problems in a realistic scenario. Additionally, the responses to both questions were missing some key concepts that were discussed during lectures but are usually not found in online resources, which contributed to a few points being taken off. In exam 2, we see a sharp decline in ChatGPT's performance, with problem 1 receiving a score of 20/75, and problem 2 getting a total of 7/25 points. This made the total grade for the exam 27/100, which is not a passing grade. Additionally, it is far lower than the human average grade for this exam, which was 71.6/100, or 71.6%. For the first question in this exam, ChatGPT struggles with understanding the question, and in turn, gives the wrong answer. The instructor's comments include that they would *"give a few mercy points"* but ultimately would fail the student as a caution to them causing any harm if they were to approach a real-world problem in a similar manner. In the second question of this exam, ChatGPT was able to understand the first part, which was a calculation. In part two, ChatGPT was unable to make a *"common sense"* assumption regarding room temperature and thus, completed the problem incorrectly.

      Another senior-level undergraduate course, EVEN 466: Sustainability and Life Cycle Analysis, was graded based on a single midterm exam. This exam is somewhat different than the other courses discussed so far with the midterm questions being completely text based with no quantitative or computational reasoning necessary. A total of 100 points were available over four questions with the point allotment as follows, 15 points for question 1, 30 points for question 2, 30 points for question 3, and 25 points for question 4. ChatGPT completed question 1 with full points (15/15) with the faculty commenting that the question was a *"gimme"* question but was impressed with how well written the answer was, mentioning that it was *"actually better written than the human median in the class"*. Question 2 when graded was found to miss an emphasis on some key concepts and texts that were used in the course. The faculty also mentioned that ChatGPT *"fumbled"* more ambiguous parts of the question and *"didn't provide the best logic to support the answer given"* but was sufficient enough to secure a 22/30 points. Question 3 demonstrates the lack of training data ChatGPT has access to where it does not analyze the specific readings from which the course is based. The faculty mentions that this answer in particular would suggest the student had not read any class readings but had taken a general ethics course at some point and secured 12/30 points. The final question, question 4, showcases how ChatGPT may be *"flawed in the way that many humans' understanding of this concept is flawed"* suggesting that it would be *"the kind of imperfect data that a non-expert would give"*

and the faculty grading this work found it *"interesting that ChatGPT's machine learning has ingested a lot of poor understanding that this type of course takes pains to disabuse students of"*. This final question earned 20/25 points, finishing with an overall total score of 69/100 points or 69% where the class average was 66/100 points or 66%. A significant change from the other senior-level course discussed here, CVEN 458 where ChatGPT did well enough to potentially perform adequately enough to move on in the course, but still slightly below what would be considered a passing score.

A graduate-level course, CVEN 664: Water Resources Planning and Management, consisted of a midterm exam that was factored into a final grade. The exam had four questions, and ChatGPT provided answers for all of them. However, the quality of the work ChatGPT submitted was quite poor and was never able to achieve the full potential points in any of the categories that were graded. Additionally, ChatGPT struggled the most with the proper notation required, validity testing, and recommendations, with each of these categories receiving only 0-2.5 points out of the possible 9-10. ChatGPT's total final grade for the midterm was 30/100, or 30%, which is not above the passing threshold. Furthermore, the average human student did far better than AI, with the average grade being 81.3/100, or 81.3%. The instructor's feedback on this rubric would also be considered very concerning, with the instructor recommending that if this was a student grade, they would organize a personal meeting to discuss student performance and the potential for dropping the course.

### 3.4 | SUMMARIZED GRADING

Four courses were graded by the faculty but not extensively commented on; one of the four was a stacked course as an undergraduate junior-level and a graduate level course, ECEN 466/776. The titles of those courses are as follows: ECEN 303: Examination Probability and Random Variables; ECEN 461: Electronic Noise; and ECEN 466/776: Information Theory, Inference and Learning Algorithms. These courses will be discussed individually, except the stacked course as a single course.

The faculty for the course ECEN 303 provided and graded ten assignments and three exams. The performance of ChatGPT on the ten assignments earned 32.25/56 possible points or 57.59% of the total possible points. The performance for the three exams was as follows: 13/20 points or 65% for Exam 1, 13.5/20 points or 67.5% for Exam 2, and 22.5/40 points or 56.25% for

Exam 3. The overall grade this faculty awarded for ChatGPT-generated responses was 60.6%, enough to secure a D in the course, although well below what would be considered a satisfactory grade for progress in the program at Anonymous University.

The faculty for the course ECEN 461 provided and graded three exams. The performance of ChatGPT's responses was very poor for this course, scoring a 33.3% for both Exam 1 and Exam 2 while scoring a 24% on the third Exam. While the collective course grade was not provided by this faculty, it would be safe to conclude that ChatGPT's performance was not adequate to pass the course.

The stacked course ECEN 466/776 was provided and graded by the same faculty as ECEN 461. In these courses, different exams were given to the undergraduate and graduate sections as appropriate for the variable difficulty between undergraduate and graduate coursework. For this course, there were two exams for both sections. The responses generated by ChatGPT for section 466 earned 23% for Exam 1 and 57% for Exam 2 while the responses for section 776 were 28% and 44%. Considering these low scores, the course would not be passed.

### 3.5 PERFORMANCE DISCUSSION

In this study the types of questions that ChatGPT performed well on were directly related to foundational levels of Bloom's Taxonomy: remembering and understanding. For the sophomore- and junior-level engineering courses, this resulted in passing grades in the courses but below the human average for the course. Faculty mentioned the solutions provided by ChatGPT were general in nature, lacking creativity in many solutions. On the other hand, senior- and graduate-level engineering courses surveyed grades earned by ChatGPT were slightly below, if not significantly below in some cases, what would be considered a passing score in the course. This suggests that while ChatGPT at the most accessible version can aid students in the development of the lowest levels of Bloom's Taxonomy, there is a gap in the ability of ChatGPT to perform adequately at these higher levels, which more accurately determine the capabilities of an engineering practice.

### 4.0 | CONCLUSIONS AND FUTURE WORK

As our world continues to develop and GAI advances, it is evident that understanding how it will ultimately impact engineering education is far from completely understood. This work was done to develop an understanding of how the GAI systems have begun in impact Texas A&M University students and faculty as well as the capabilities of ChatGPT, the most prominent GAI system at the time this work began, to complete coursework provided by engineering faculty at Texas A&M University. Since the original release of ChatGPT, several other companies have developed GAI systems that are comparable, and OpenAI now offers an improved more dynamic GAI system on a subscription-based model. Texas A&M University has also purchased a license for access to the Microsoft Copilot GAI system accessible for faculty and staff at the university. This technology will clearly become integrated into many facets of our lives and will impact the way we conceptualize our work in the future.

Our preliminary findings highlight the need for continued adaptation to the changing role of generative AI in education. Building on the DANCE model, future research will explore refining detailed teacher development programs tailored to integrate emerging technologies such as ChatGPT (Shryock et al., 2024). We plan to investigate the longitudinal impact of these technologies on academic perceptions and learning outcomes, refining the DANCE model to help faculty address the challenges and opportunities presented by generative AI systems.

Given that consideration the researchers have decided to develop a model to understand the ways in which disruptions enact change just as ChatGPT has influenced change within the higher education community. A second iteration of a survey has also been developed and will be distributed to the Texas A&M University community during the Spring of 2024 semester. This will inform the authors as to how the integration of GAI has impacted the perceptions of faculty, staff, and students in a longitudinal way as well as allow the researchers to begin the validation of the survey instruments. More advanced statistical methods are planned to examine this data including a more thoroughly developed approach to network analysis to understand the underpinning impacts of factors, such as department, position, and role of faculty in the university as it pertains to their perceptions of GAI.

## 5.0 | ACKNOWLEDGMENTS

We would like to acknowledge the assistance of the faculty who contributed to this work by sharing their course materials and assessments of ChatGPT's performance on that course work, Dr. Kelly Brumbelow. Dr. Laszlo Kish, and J.F. Chamberland. Work with Dr. Jonan

Donaldson in the development of the network analysis was incredibly insightful to tie together the different respondent groups who shared different overall perceptions of ChatGPT; without his help that portion of this work would not have been possible. We would also like to acknowledge the foundational work done by two undergraduate students who worked with the research team early in the assessment of the course materials through the usage of ChatGPT, Alexis Lizarraga and Laksha Arora.

**6.0 | ABOUT THE AUTHORS**

Lance White: Lance White is a PhD candidate at Texas A&M University in the Multidisciplinary Engineering department. He is also a Lecturer for First-Year Engineering in Engineering Academic and Student Affairs at Texas A&M University. His research interests include curricular evolution, diversity, equity, and inclusion, faculty development, and organizational change.

Trini Balart: Trini Balart is a Multidisciplinary Engineering PhD student at Texas A&M University, with a focus in engineering education. Her main research interests include the improvement of engineering students' learning, innovative ways of teaching and learning, and how artificial intelligence can be used in education in a creative and ethical way.

Sara Amani: Sara Amani is a doctoral student at Texas A&M University. She completed her Bachelors of Science in chemical engineering from Texas A&M University at Qatar. She is currently pursuing her PhD in Multidisciplinary Engineering with a focus in engineering education. Her research interests include mental health in engineering education and women in engineering.

Dr. Kristi J. Shryock: Kristi J. Shryock, Associate Professor in Multidisciplinary Engineering at Texas A&M University, works to form the engineer of the future through the network of transformational strategies she has developed, which include informing early, addressing preparation for success, increasing diversity, establishing engineering identity, developing personalized educational opportunities, and enhancing professional skills.

Dr. Karan L. Watson: Karan L. Watson, Regents Senior Professor of Electrical and Computer Engineering and Provost Emeritus, is a fellow of the Institute of Electrical and Electronic Engineers, the American Society for Engineering Education, and the Accreditation

Board for Engineering and Technology. Among her awards is the ASEE Lifetime Achievement in Engineering Education.

# References

Amani, S., White, L., Balart, T., Arora, L., Shryock, D. K. J., Brumbelow, D. K., & Watson, D. K. L. (2023). Generative AI Perceptions: A Survey to Measure the Perceptions of Faculty, Staff, and Students on Generative AI Tools in Academia. *arXiv preprint arXiv:2304.14415*.

Girvan, M., & Newman, M. E. (2002). Community structure in social and biological networks. *Proceedings of the national academy of sciences*, *99*(12), 7821-7826.

Shryock, K., Watson, K., White, L., & Balart, T. (2024). Developing a Model to Assist Faculty with Taming the Next Disruptive Boogeyman. *SSRN: 4699941.* [InPress: Applied Computing eJournal, May 2024]

Wilson, L. O. (2016). Anderson and Krathwohl–Bloom's taxonomy revised. *Understanding the New Version of Bloom's Taxonomy*.